# Ultrathin liquid crystal metasurface for on-demand terahertz beam forming over 110° field-of-view

Chen Chen[1,2], Sai Chen[1,2,3], Yibo Ni[1], Yibao Xu[1], Yuanmu Yang[1*]

[1]State Key Laboratory for Precision Measurement Technology and Instruments, Department of Precision Instrument, Tsinghua University, Beijing 100084, China.

[2]These authors contributed equally: Chen Chen, Sai Chen.

[3]Current address: School of Electronic Information Engineering, Beihang University, Beijing 100191, China.

*Corresponding author: ymyang@tsinghua.edu.cn

## Abstract

**The terahertz spectral region, which bridges between electronics and optics, is poised to play an important role in the development of transformative wireless communication and imaging systems with unprecedented functionality. Currently, a major challenge in terahertz technology is to develop high-performance terahertz beam-forming devices that can dynamically shape the terahertz radiation in a flexible manner. Existing terahertz beam-forming devices have limited coding bits, field-of-view, and beam gain. Here, we experimentally demonstrate a reconfigurable liquid crystal-integrated terahertz metasurface, with each metasurface unit cell being independently addressable. The metasurface has a 260° continuous phase tuning range with a device thickness of only 1% of the free-space wavelength. We show that the terahertz wave diffracted from the metasurface can be steered towards a wide range of directions, covering a record-large 110° field-of-view with a peak gain of 25 dBi. The metasurface also features a low power consumption and sub-second switching time. Furthermore, we demonstrate the formation of multiple terahertz beams, with the direction of each beam and the power ratio between beams adjustable on demand. The proposed liquid crystal metasurface possesses compelling prospects for future terahertz communication and imaging applications.**

High-speed wireless communication is of critical importance under the unprecedented rapid explosion of digital content. From Shannon's formula, the capacity of a communication channel is directly proportional to the carrier bandwidth. Consequently, terahertz communication with an ultra-wide bandwidth has been considered promising for constructing the next-generation high-speed wireless communication system[1-4].

Compared to conventional microwave-based systems, a key challenge of terahertz communication systems is the high atmospheric absorption and free space path loss of terahertz radiation. To compensate for this high path loss and to extend the link coverage, a new type of system component, reconfigurable intelligent surface (RIS), has been proposed to reshape the communication environment[5-9]. RIS, sometimes also referred to as reconfigurable metasurface[10-12], is a programmable surface structure that can be used to manipulate the reflection of electromagnetic waves in a flexible manner. For instance, when reflected waves interfere constructively in a single direction, a highly directional beam is formed to overcome the severe path loss[13-15]. Using the reconfigurability of the metasurface to re-orient the directional beam, one can extend the field-of-view (FOV) of the communication link. Furthermore, a single metasurface can serve multiple users if multiple independent beams can be simultaneously formed, which can drastically simplify the communication system, as schematically illustrated in Fig. 1.

The use of RIS has been widely exploited at microwave frequencies, mostly using diodes as the active medium[16-18]. However, an array of diodes may consume a large amount of power, and their operation frequency is restricted by the parasitic loss. To realize a reconfigurable metasurface for operation at the terahertz spectral regime[19], one may incorporate alternative active mediums, such as silicon[20,21] or III-V semiconductor[22-24] into the metasurface design. Nonetheless, the device fabrication typically requires advanced semiconductor processing tools or foundries, with limited antenna aperture size and angular resolution for beam forming. MEMS devices[25-27], graphene[28-31], and phase change matrials[32-34] have also been proposed for the active tuning of the THz beam, yet pixelated control of the meta-atoms via electrical biasing remains a major challenge.

On the other hand, liquid crystal is a relatively mature material that is widely used for tuning electromagnetic waves spanning from microwave to visible frequencies[35]. The orientation of liquid crystal molecule, and consequently its refractive index, can be controlled by an external electric field straightforwardly. The incorporation of meta-structures in a liquid crystal cell can significantly enhance the light-matter interaction, leading to a relaxed requirement on liquid crystal layer thickness for wide-range phase tuning, resulting in more uniform liquid crystal molecular alignment, lower control voltage, and increased switching speed. Liquid crystal-based devices typically have a low power consumption. Moreover, it has been proven that large-area liquid crystal panels can be readily mass-produced at a low cost.

To date, there have been several demonstrations of dynamic terahertz beam-forming devices using liquid crystal-integrated metasurfaces[36-44]. Some of which adopt the coding metasurface concept[37,39-41,44] having only 1 coding bit, with restricted FOV and oftentimes unwanted sidelobes. For other demonstrations that allow continuous phase tuning, their applications are currently hindered due to limited beam gain and FOV due to transmitter blockage. Furthermore, previous studies on THz liquid crystal metasurfaces[36-44] have a relatively thick liquid crystal layer, resulting in slow switching speeds. Terahertz beam-forming device with combined high gain, low sidelobe, large FOV, and high switching speed remains a critical challenge for practical applications.

In this work, we experimentally demonstrate a liquid crystal-integrated metasurface for on-demand terahertz beam-forming. We identify a metal-insulator-metal meta-atom design with narrow inter-unit gaps that allows a continuous 260° phase tuning range with a liquid crystal layer thickness of only 1% of the incident wavelength, leading to a sub-second switching time. With the beam incident at a near-grazing angle of 75°, the metasurface features a record-wide FOV of 110° that is free from transmitter blockage. With 80 meta-atoms independently controlled by an optimized applied voltage pattern, the beam gain at the operation frequency of 0.29 THz can reach up to 25 dBi with sidelobes below -10 dB and the peak steering efficiency reaching 30%. Furthermore, leveraging the flexible and continuous phase tuning capability, we demonstrate on-demand control of the direction and power ratio of multiple formed beams, which is critical for simultaneously providing communication service to multiple users and for performing system networking.

**Metasurface design, fabrication, and characterization**

For the metasurface design, the widely adopted metal-insulator-metal configuration[36-45] is chosen due to its simplicity and compatibility with the mature liquid crystal display assembly line. Two layers of gold serve as the ground plane and the individually addressable electrode, respectively, with the liquid crystal serving as the insulator. A cross-sectional view of two adjacent unit cells is illustrated in Fig. 2a, with a photograph of the whole metasurface shown in Fig. 2b. From the effective circuit analysis, to reach the resonant condition of a metal-insulator-metal structure with a minimum insulator thickness, one can minimize the gap width between adjacent unit cells (details in Supplementary Section 1). Here, the gap width is set to 5 μm, only restricted by the available photolithography tool. The resulting liquid crystal layer thickness is 12 μm, which is only about 1% of the free-space operation wavelength (1034 μm at 0.29 THz). The liquid crystal mixture used here has a large birefringence and low absorption ($n_o$ = 1.66, $n_e$ = 1.94, $k_o$ = 0.025, $k_e$ = 0.05) at 0.29 THz.

To avoid the transmitter (Tx) blockage, the metasurface is designed to operate under a glazing incident angle of 75°. The reflectance and phase spectrum of the metasurface with *p*-polarized incidence are shown in Fig. 2c. A clear resonance around 0.29 THz is observed from the reflectance spectrum. The resonance frequency is redshifted by 27 GHz with the liquid crystal director angle increasing from 0° to 90°. With 0.29 THz selected as the operation frequency ($f_{OP}$), we obtain the widest continuous phase tuning range of 280°, accompanied by a large reflectance tuning, as shown in Fig. 2d. The phasor diagram showing the corresponding variation of the complex reflection coefficients is depicted in Fig. 2e.

The fabrication of the metasurface and its assembly with a liquid crystal cell follows a rather standard protocol (details in Supplementary Section 2). Subsequently, the metasurface is fixed onto a custom-made holder, with 80 unit cells individually wire-bonded to a printed circuit board. The voltage-dependent reflectance and phase spectrum of the metasurface are measured with an angle-resolved terahertz time-domain spectroscopy (THz-TDS) system (details in Methods). An AC square wave with the peak-to-peak voltage ($V_{PP}$) swept from 0 V to 35 V is uniformly applied to all unit cells. Using the reflectance spectrum of a gold plate as a reference, the reflectance and phase tuning spectrum of the metasurface are measured and shown in Fig. 2f. Subsequently, the reflectance and phase tuning curves at 0.29 THz are extracted and shown in Fig. 2g, together with the phasor diagram of complex reflection coefficients shown in Fig. 2h. The metasurface is capable of performing continuous phase tuning up to 260°, which closely matches the simulation results. Moreover, we verify the phase measurement uncertainty of the THz-TDS system to be about 2°, such that continuous tuning of 260° corresponds to 7 coding bits.

**Terahertz beam forming towards a single direction**

To form the terahertz beam to a specified target angle $\theta_{TA}$, the desired phase at each unit cell can be calculated from the grating equation $\varphi(m) = -mk_{OP}p\ (\sin\theta_{Tx}+\sin\theta_{TA})$, where $m$ is the unit index in the array, $k_{OP} = 2\pi f_{OP}/c$ is the wavenumber at the operation frequency, $c$ is the speed of light in free space, $p$ is the grating constant, and $\theta_{Tx}$ is the angle of the incident beam.

The schematic of the experimental setup for THz beam forming is illustrated in Fig. 3a, along with a photograph shown in Fig. 3b (details in Methods). According to the measured continuous phase tuning curve, one can convert the desired phase at each metasurface unit cell to an applied voltage. Due to the limited phase tuning range of the metasurface, we use the replace-half-half strategy[46] to replace the unavailable phase values with 0° or 260°, which further translates to a control voltage of 0 V or 35 V. Converted theoretical voltage patterns for target angle $\theta_{TA} = 50°, 0°$, and -50°, respectively, are shown in Fig. 3c.

By fixing the Tx at $\theta_{Tx} = -75°$, 12 beam-forming target angles are selected at a 10° interval within a total FOV of 110° from 60° to -50°. After applying the theoretical voltage pattern for each target angle, the resulting measured gain patterns are shown in Fig. 3d, which deviate largely from the theoretical predictions. The degraded performance is attributed to the mutual coupling between metasurface unit cells. The voltage pattern in Fig. 3c is obtained with local phase approximation[47], where the coupling between neighboring meta-atoms is neglected. However, such an approximation is invalid in our metasurface with a highly nonlocal response.

To further enhance the beam-forming performance, the voltage pattern is optimized according to the Stochastic Parallel Gradient Descent (SPGD) algorithm[48] using the beam amplitude measured by the receiver at the target angle as the Figure-of-Merit (*FoM*) (details in Supplementary Section 3). The optimized voltage patterns for $\theta_{TA}$ = 50°, 0°, and -50°, respectively, are shown in Fig. 3e. The resulting optimized gain patterns for $\theta_{TA}$ ranging from 60° to -50° are shown in Fig. 3f.

The gain, steering efficiency, and primary-side-lobe-ratio (PSLR) at each $\theta_{TA}$ are tabulated in Supplementary Table S1, showing a maximum gain of 25 dBi, a maximum steering efficiency of 30%, and a maximum PSLR of 12 at $\theta_{TA}$ = 60°. Notice the gain is non-uniform across the entire FOV, which is attributed to the non-uniform radiation pattern of the metasurface unit cell (details in Supplementary Section 4). Bandwidth and beamwidth at each $\theta_{TA}$ are also extracted from measurements (details in Supplementary Section 5). Moreover, we show that the beam formed by the metasurface can cover the whole FOV in an even finer angular resolution of 3°, comparable to the average beamwidth, with further optimization (details in Supplementary Section 6).

**Terahertz beams forming towards multiple directions**

In daily communication scenarios, serving one user per metasurface is not cost-effective due to the ever-increasing density of digital terminals. A more realistic scenario involves serving multiple users per metasurface. Temporal multiplexing is required if the metasurface can only form one beam at a time, which is very demanding on the switching speed of the metasurface. The advanced functionality of individually forming multiple beams can relax this speed requirement. In this way, the functionality of beam splitting, beam steering, and beam power adjustment can be integrated into a single metasurface, simplifying the deployment of beam-forming metasurfaces into communication networks. Additionally, multiple formed beams can enhance the physical layer security by suppressing signal strength at eavesdroppers[49], reduce the beam training overhead in identifying user directions[50], and eventually realize more versatile communication networks.

To demonstrate multiple-beam-forming, a modified SPGD optimization process is deployed, where *FoM* is measured at each target angle and summed up with corresponding weights. Eventually, optimized voltage patterns for different combinations of target angles are obtained, with the corresponding gain patterns measured and shown in Figs. 4a-b. In particular, Fig. 4a demonstrates the metasurface's ability to form two beams with 10° separation across the 110° FOV. Fig. 4b shows the metasurface's ability to split one incident beam into three or four beams. Since multiple beams can be formed in desired directions by switching between these voltage patterns, two or more users can move freely within the wide FOV of the metasurface, without being disconnected from the network. To further improve the uniformity of gain across the entire FOV or to selectively increase the gain towards a certain direction, one may adjust the corresponding weights in the *FoM* (details in Supplementary Section 7). In addition, the metasurface can operate under a wide range of incident angles, from $\theta_{Tx}$ = -75° to 75°. As shown in Fig. 4c, the beam is formed at $\theta_{TA}$ = 40° under various incident angles from -75° to 0°, after applying optimized voltage patterns at the corresponding incident angle. This feature is highly desirable in facilitating system networking (details in Supplementary Section 8).

**Discussion and Conclusion**

We verify the long-term reliability of the fabricated metasurface as we observe the phase tuning curves of the metasurface remain consistent after 100,000 voltage upload operations. The power consumption of the 80-channel metasurface of 2×2 cm$^2$ clear aperture is measured to be only 5.6 mW (details in Supplementary Section 9). The switching speed of the metasurface is measured to be 1.6 Hz (details in Supplementary Section 10), which may be further improved by further reducing the liquid crystal layer thickness or by adopting liquid crystal molecules with reduced viscosity. To further simplify the device operation, we show that the system can also work with a binary applied voltage with moderately compromised beam forming performance (details in Supplementary Section 11). Although reconfigurable one-dimensional beam-forming is experimentally demonstrated here, two-dimensional beam-forming covering the whole hemisphere is possible through a similar design concept by extending the metasurface into a two-dimensional array.

To summarize, here we experimentally demonstrate on-demand terahertz beam-forming using a liquid crystal-integrated reconfigurable metasurface. Using a unit cell design featuring a metal-insulator-metal structure with narrow in-plane gaps, we achieve a 260° wide phase tuning range with an ultrathin liquid crystal layer. By applying optimized voltage patterns to an array of 80 individually controlled unit cells, a maximum beam-forming gain of 25 dBi, and a maximum steering efficiency of 30% are achieved within a record-wide FOV of 110° without transmitter blockage. The reconfigurable metasurface further features a small beam-forming angular resolution and a wide range of acceptable incident angles. In addition, the metasurface is capable of forming multiple beams with an adjustable gain ratio, which is critical in providing simultaneous communication access to multiple users and in system networking. The proposed beam-forming metasurface may also be adopted for terahertz imaging. Therefore, it may pave the way for a highly integrated system for simultaneous terahertz sensing and communication.

**Method**

**Simulation set-up**

We use ANSYS Lumerical FDTD for the simulation of the metasurface response. For the unit cell simulation, with an oblique incidence, the Bloch boundary condition is applied along the *x*-direction. The reflectance spectrum is obtained by sweeping the source frequency. Liquid crystal is modeled as an anisotropic material with material properties listed in the main text, with the director angle varied in the *x-z* plane from 0° to 90°. To approximate the refractive index change with an applied voltage, liquid crystal molecules inside gap regions are assumed to be fixed while molecules below the top metal patches are uniformly tuned by the applied voltage.

**Experimental set-up**

In the experimental set-up, photoconductive antennae in a fiber-based THz Time Domain Spectroscopy system (THz-TDS, Terahertz Photonics Co. Ltd.) are used as the transmitter (Tx) and the receiver (Rx) of THz waves. Tx and Rx are secured on two arms of a double-axes concentric rotation stage (OptoSigma OSMS-120YAW-W), both aligned to the stage's rotational center. The metasurface holder is fixed at the center of the rotation stage. Relative angles of the Tx and Rx with respect to the metasurface can be independently adjusted with the rotation stage. Following the angle definition as depicted in Fig. 3a, Tx is located at $\theta_{Tx}$ = -75° and the angle of Rx ($\theta_{Rx}$) can vary within a range from -55° to 90°, limited by the blockage from the Tx. THz radiation from the Tx is *p*-polarized, with the electric field vector within the rotational plane. The control signal to each unit cell is a 3 kHz AC square wave generated by a field-programmable gate array (FPGA) board. The peak-to-peak voltage ($V_{PP}$) of the control signal applied to each unit cell can be individually tuned in a range from 0 V to 35 V. The FPGA board is supplied with a DC power supply, connected to the printed circuit board by wire bundles, and controlled by a computer. The reflected terahertz wave from a gold plate the same size as the metasurface is used as the reference for reflectance, phase, gain, and efficiency measurements.

## Acknowledgments

This work was supported by the National Natural Science Foundation of China (62135008, 61975251) and by the Guoqiang Institute, Tsinghua University. The authors thank Yuehong Xu at Terahertz Photonics Co. Ltd. for her help with the THz-TDS system and thank Shuzhe Li at Chin Instrument Inc. for his help with the FPGA control board.

## Author contributions

Y.Y. and S.C. conceived the idea; C.C., S.C., and Y.N. performed the simulations; C.C. and S.C. prepared the sample and carried out experimental measurements; C.C. and Y.X. implemented the optimization algorithms; all authors contributed to analyzing the data and writing the manuscript; Y.Y. supervised the project.

## Competing financial interests

The authors declare no competing financial interests.

# Figures

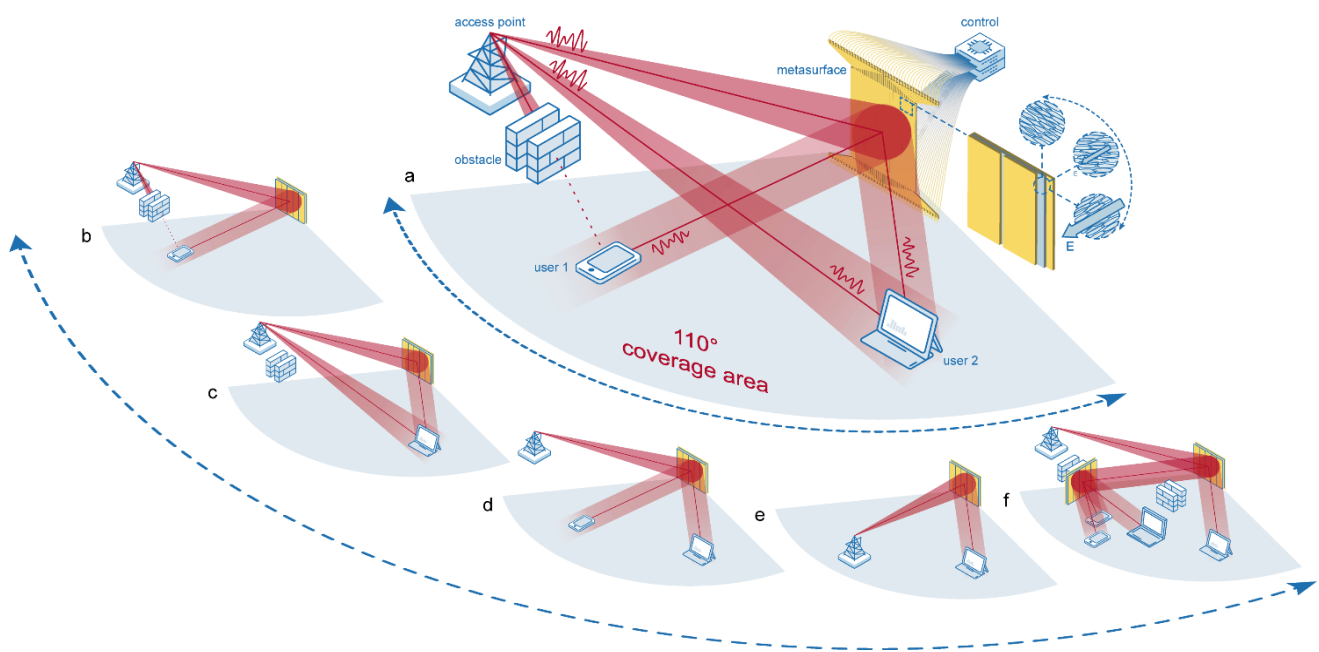

**Figure 1 | Schematic illustration of using the liquid crystal metasurface for on-demand wide FOV terahertz beam-forming. a,** Schematic of the liquid crystal-integrated metasurface, controlled by electrical voltages, which can form the THz beam over a 110° field-of-view. **b-f,** Potential applications of the reconfigurable liquid crystal-integrated metasurface, including **(b)** bypassing obstacles via an alternative signal path, **(c)** strengthening the wireless signal via multiple paths, **(d)** simultaneously serving multiple users with multiple formed beams, **(e)** redirecting the wireless signal with a varied incident beam direction, **(f)** performing system networking to adapt to complex communication scenes.

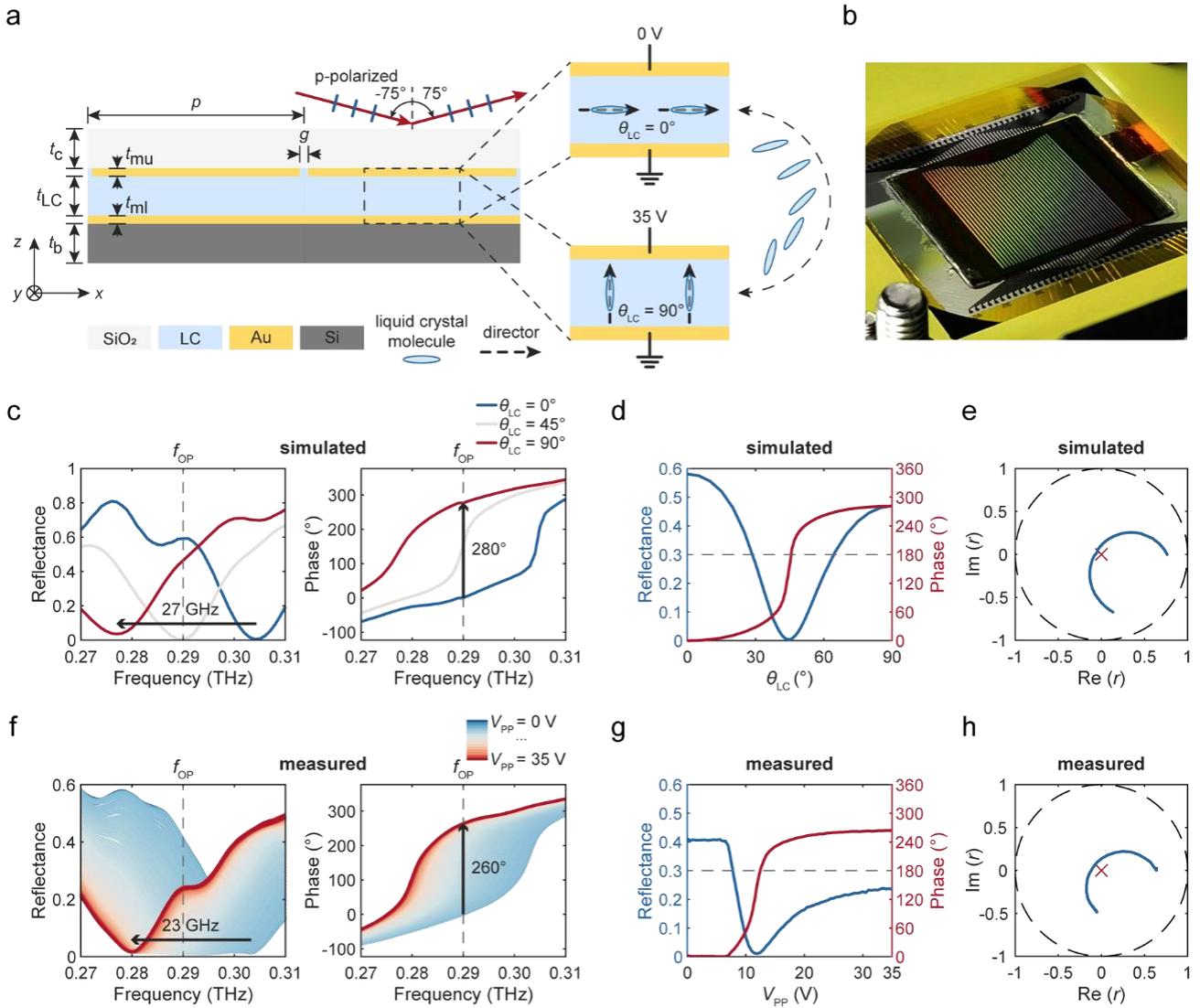

**Figure 2 | Metasurface design and characterization. a**, Cross-sectional view of the metasurface unit cell (not shown to scale), with $p$ = 240 μm, $g$ = 5 μm, $t_c$ = 1000 μm, $t_{LC}$ = 12 μm, $t_b$ = 500 μm, $t_{mu}$ = 0.2 μm, and $t_{ml}$ = 0.1 μm. The $p$-polarized THz wave incidents at an angle of 75°. **b**, Photograph of the fabricated metasurface fixed and wire-bonded to the holder. The aperture size is 2×2 cm². **c**, Simulated metasurface reflectance and phase spectrum with $\theta_{LC}$ = 0°, 45°, and 90°, respectively. **d**, Simulated reflectance and phase tuning curves at $f_{OP}$ = 0.29 THz (labeled by the dashed line in **c**), with the liquid crystal director angle varying from 0° to 90°. **e**, Corresponding phasor diagram of simulated reflection coefficients. **f**, Measured reflectance and phase spectrum as a function of peak-to-peak voltage ($V_{PP}$) uniformly applied to all unit cells. **g**, Measured reflectance and phase tuning curves at $f_{OP}$ = 0.29 THz (labeled by the dashed line in **f**). **h**, Corresponding phasor diagram of measured reflection coefficients.

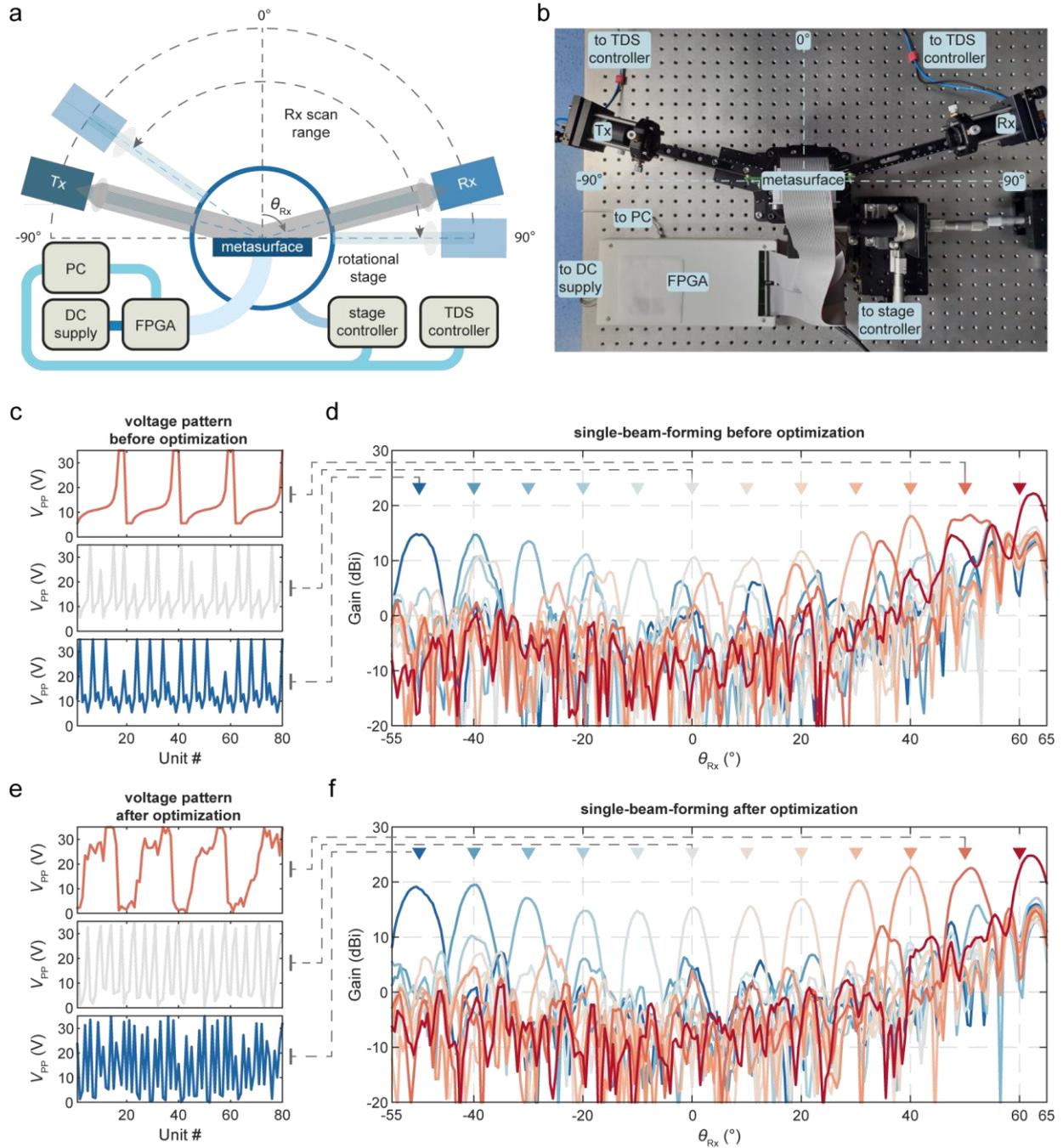

**Figure 3 | Experimental set-up and single-beam-forming performance of the metasurface. a-b**, Schematic illustration (**a**) and photograph (**b**) of the experimental set-up. **c**, Theoretical voltage patterns without considering the mutual coupling between adjacent unit cells for target angles $\theta_{TA}$ = 50°, 0°, and -50°, respectively. **d**, Measured gain patterns after applying theoretical voltage patterns for 12 target angles from 60° to -50°, at a 10° interval. **e**, Optimized voltage patterns for $\theta_{TA}$ = 50°, 0°, and -50°, respectively. **f**, Measured gain patterns after applying optimized voltage patterns for 12 target angles from 60° to -50°, at a 10° interval.

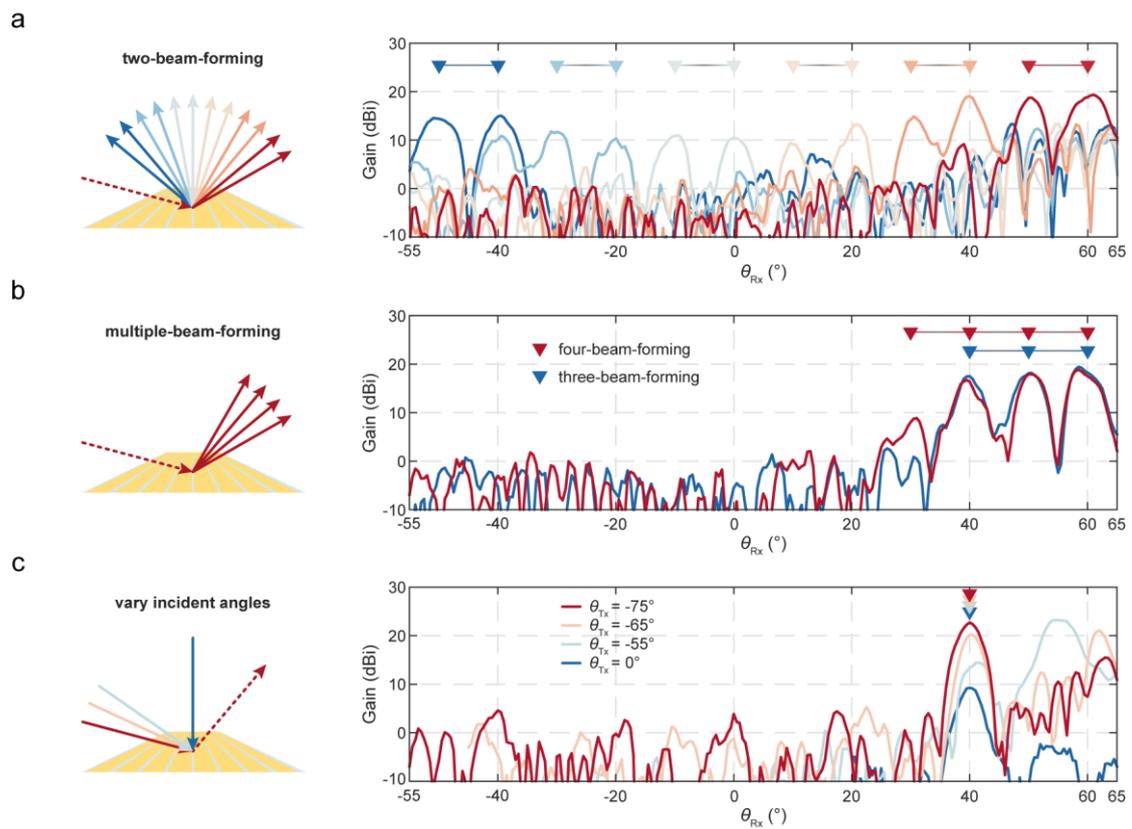

**Figure 4 | Multiple-beam-forming performance and operation with various incident angles. a**, Measured two-beam-forming results with 10° separation across the 110° FOV. **b**, Measured three-beam-forming and four-beam-forming results. **c**, Measured beam-forming results with $\theta_{TA} = 40°$ and with incident angle $\theta_{Tx}$ = -75°, -65°, -55°, and 0°, respectively, after applying optimized voltage patterns at the corresponding incident angle.

# Supplementary Information: Ultrathin liquid crystal metasurface for on-demand terahertz beam forming over 110° field-of-view

Chen Chen[1,2], Sai Chen[1,2,3], Yibo Ni[1], Yibao Xu[1], Yuanmu Yang[1*]

[1]State Key Laboratory for Precision Measurement Technology and Instruments, Department of Precision Instrument, Tsinghua University, Beijing 100084, China.
[2]These authors contributed equally: Chen Chen, Sai Chen.
[3]Current address: School of Electronic Information Engineering, Beihang University, Beijing 100191, China.
*Corresponding author: ymyang@tsinghua.edu.cn

**Section 1: Resonance mode and equivalent circuit model**

The field distribution at the resonance frequency, as shown in Fig. S1a, reveals the magnetic dipole nature of the resonant mode. With the liquid crystal molecules initially aligned along the *x*-axis, the applied voltage can significantly alter its refractive index along the *z*-axis, resulting in a large resonance frequency redshift.

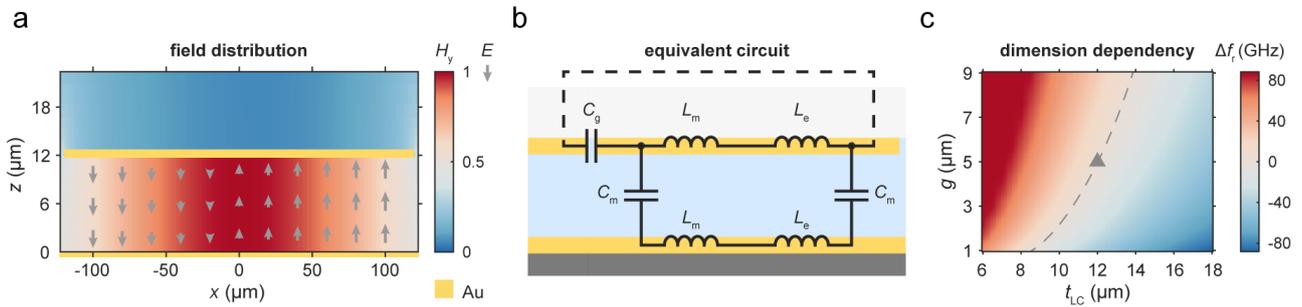

**Figure S1 | Field distribution on resonance, equivalent circuit model, and dependency of resonance frequency on structural dimensions. a**, Cross-sectional view (not shown to scale) of the *y*-component of magnetic field ($H_y$) and the vectorial electric field ($E$) at resonance frequency $f_r = 0.29$ THz. **b**, Equivalent circuit model for analyzing the resonance frequency of the metal-insulator-metal resonator. **c**, Simulated resonance frequency variation $\Delta f_r$ from $f_r$ of the proposed meta-atom (triangular mark) as a function of the liquid crystal layer thickness $t_{LC}$ and gap width $g$. The equal-frequency line (dashed line) where $\Delta f_r = 0$ shows that while keeping the resonance frequency fixed, to reduce $t_{LC}$, $g$ needs to be reduced accordingly.

Moreover, an equivalent circuit model is adopted to verify the resonance frequency's dependence on the meta-atom's structural dimensions. As observed from the cross-section of the metasurface, the structure can be described as a linear circuit consisting of capacitors and inductors (Fig. S1b), with the value of each circuit element determined by structural dimensions and material properties[1,2]. The parallel plate capacitance $C_m = 0.5\varepsilon_{LC,z}wl/t_{LC}$, where $\varepsilon_{LC,z}$ is the liquid crystal permittivity in the $z$-direction, $w$ is the width of the metal patch in the $x$-direction, $l$ is the length of the metal patch in the $y$-direction, and $t_{LC}$ is the thickness of the liquid crystal layer in the $z$-direction. The gap capacitance $C_g = 2\varepsilon_g \mathrm{acosh}((2w+g)/g)l/\pi$, where $\varepsilon_g$ is the permittivity in the gap, which is assumed to be equal to the liquid crystal permittivity along its optical axis $\varepsilon_{/\!/}$, and stays constant with applied voltages, and $g$ is the gap width. The parallel plate inductance $L_m = 0.5\mu_0 w t_{LC}/l$, where $\mu_0$ is the vacuum permeability. The kinetic inductance $L_e = -w/(\varepsilon_0 \varepsilon_m' t_{m,\mathrm{eff}} l \omega^2)$, where $\varepsilon_0$ is the vacuum permittivity, $\varepsilon_m'$ is the real part of gold's permittivity, obtained from the Drude model fit[3], $t_{m,\mathrm{eff}}$ is the effective thickness of the metal, defined as the minimum between metal thickness and metal skin depth, and $\omega = 2\pi f$ is the angular frequency. The total impedance of the circuit is

$$Z(\omega) = i\omega \left[ \frac{L_m + L_e}{1 - (L_m + L_e)\omega^2 C_g} - \frac{2}{\omega^2 C_m} + L_m + L_e \right] \tag{S1}$$

From the equivalent principle, the resonance frequency of the circuit reflects that of the metasurface. At the resonance frequency $f_r$, the total circuit reactance $\mathrm{Im}(Z(2\pi f_r)) = 0$. The dependency of resonance frequency on spacer thickness $t_{LC}$ and gap width $g$ is shown in Fig. S1c. For a liquid-crystal-integrated metasurface, the thickness of the liquid crystal layer $t_{LC}$ needs to be minimized to achieve an increased switching speed. To keep the resonance at a specified frequency, the gap width $g$ needs to be scaled down accordingly. Thus, the proposed metasurface has very small gap openings between top metal patches, forming a metal grating with a large duty cycle ($w/(w+g) \approx 98\%$).

**Section 2: Fabrication and assembly of the liquid crystal metasurface**

The metasurface mainly comprises three layers, the top metal patch, the metal ground plane, and the middle liquid crystal layer. The metal patch is formed by the standard photolithography and lift-off process. The metal ground plane is a commercially purchased Au-on-silicon wafer, diced to fit the size of the metasurface.

The liquid crystal cell formation process is shown in Fig. S2a. After both metallic layers are cleaned, a layer of SD1 is spin-coated on both metallic surfaces (Step 1). Linearly polarized UV light exposed on the SD1 films defines the liquid crystal's alignment direction (Step 2). SiO$_2$ spheres are then sprayed onto the patterned metal (Step 3). Subsequently, the metal ground plane is placed on the sprayed spacers, with the distance between the two metallic layers defined by the diameter of the SiO$_2$ spheres (Step 4). Next, capillary filling of the liquid crystal molecules (BDLC) is performed at an elevated temperature (Step 5). After the filling, four edges of the metasurface are sealed with glue, forming an enclosed cell (Step 6).

After the liquid crystal cell formation process, extended pads from top metal patches are wire-bonded to the printed circuit board. A portion of the wire-bond region is shown in Fig. S2b.

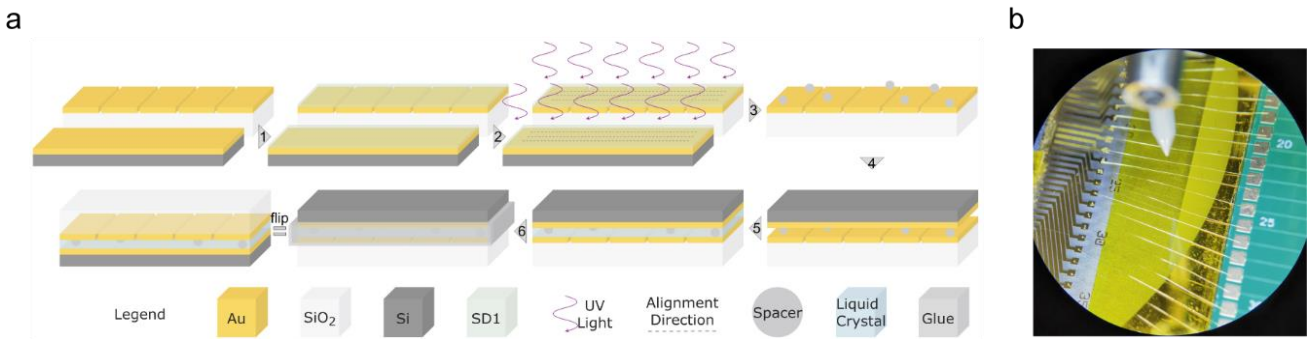

**Figure S2 | Fabrication and assembly process of the metasurface. a**, Liquid crystal cell formation process. **b**, Picture of the wire-bond region.

### Section 3: Optimization algorithm for metasurface with mutual coupling

In the proposed metasurface, the mutual coupling between unit cells is high, as indicated by the high electric field inside the gap region shown in Fig. S3a. A gradient descend-based optimization algorithm, the Stochastic Parallel Gradient Descent (SPGD) algorithm, is implemented to search for the optimum voltage pattern from a known starting point. The optimization process can also account for fabrication imperfections between unit cells and the channel variation from FPGA outputs.

The flowchart of the SPGD algorithm is shown in Fig. S3b, with the main procedures described as follows. The Figure-of-Merit (*FoM*) in the SPGD algorithm is defined as the formed beam amplitude at the target angle. During each iteration, the first *FoM* is measured after applying a random perturbation *dV* to the voltage pattern. After the reversed perturbation -*dV* is applied to the voltage pattern, another *FoM* is measured. Through the update formula shown in Fig. S3b, an updated voltage pattern is determined with *dV* and two measured *FoM*s. The evolution of the *FoM* for $\theta_{TA} = 50°$ during 800 iterations is shown as an example in Fig. S3c.

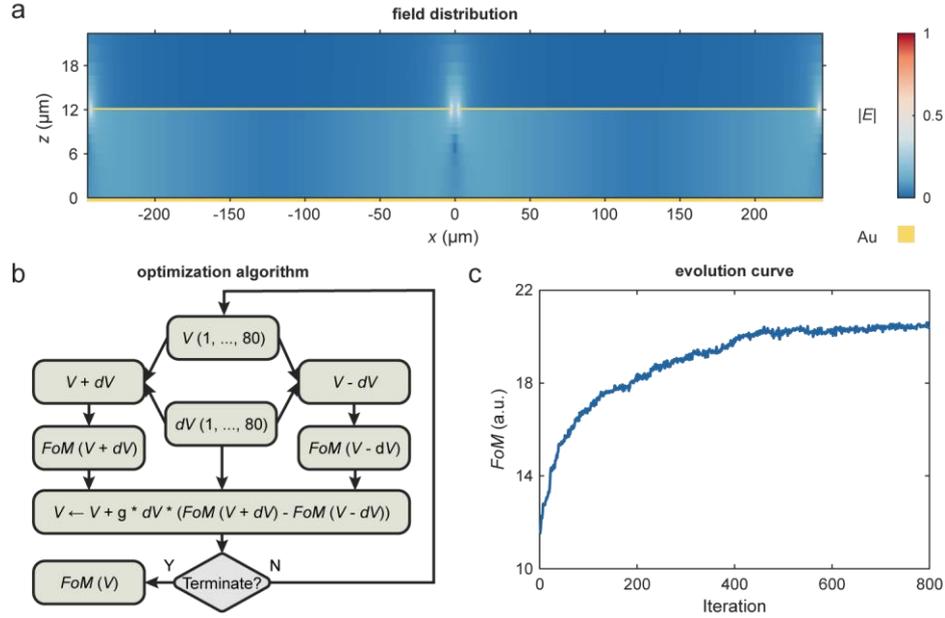

**Figure S3 | Optimization algorithm for high mutual coupling metasurface, and Figure of Merit evolution curve. a**, Cross-sectional view (not shown to scale) of the normalized electric field strength (|*E*|) at $f_{OP} = 0.29$ THz, showing high electric field strength in the gap region between two adjacent unit cells, indicating high mutual coupling. **b**, Flow chart of the Stochastic Parallel Gradient Descend (SPGD) algorithm used in optimizing the voltage pattern, where *V* is the voltage pattern, *dV* is a random voltage perturbation, and g is a scaling constant. **c**, Figure of Merit (*FoM*) evolution curve for $\theta_{TA} = 50°$ during the iterative optimization process.

**Table S1 | Single-beam-forming performance at various target angles.**

| Target Angle | 60° | 50° | 40° | 30° | 20° | 10° | 0° | -10° | -20° | -30° | -40° | -50° |
|---|---|---|---|---|---|---|---|---|---|---|---|---|
| Gain (dBi) | 24.8 | 22.5 | 22.7 | 20.2 | 16.8 | 15.6 | 15.4 | 14.8 | 14.8 | 17.1 | 19.5 | 19.1 |
| Steering Efficiency (%) | 29.5 | 17.5 | 18.2 | 10.3 | 4.7 | 3.5 | 3.4 | 2.9 | 2.9 | 5.0 | 8.7 | 8.0 |
| PSLR | 11.8 | 5.5 | 5.3 | 4.5 | 1.5 | 1.5 | 1.5 | 1.4 | 0.6 | 1.4 | 2.4 | 2.1 |
| Bandwidth (GHz) | 21 | 15 | 17 | 18 | 16 | 15 | 13 | 13 | 10 | 10 | 9 | 9 |
| Beamwidth (°) | 4.5 | 5.0 | 3.7 | 3.7 | 4.1 | 3.2 | 3.3 | 3.5 | 3.9 | 3.7 | 4.1 | 5.2 |

**Section 4: Limitation from the single unit cell radiation pattern**

As can be observed in Table S1, the beam-forming performance degrades with decreasing $\theta_{TA}$, with a minimum around $\theta_{TA} = -20°$, where the gain drops to 15 dBi and the PSLR is below 1. This observation is attributed to the radiation pattern from a single unit cell, also known as the unit factor, which inevitably limits the steering efficiency to certain target angles.

By putting a dipole source inside one unit cell of an array structure and recording the far-field radiation pattern, the single unit cell radiation pattern is simulated and plotted in Fig. S4a, showing nulls in both $|\theta_{Rx}| = 0°$ and $|\theta_{Rx}| = 90°$, and peak around $|\theta_{Rx}| = 60°$. This pattern imposes an envelope on the gain, steering efficiency, and PSLR at various target angles, shown in Fig. S4b, which matches curves of normalized gain, steering efficiency, and PSLR plotted in the same figure.

Analytically, this radiation pattern resembles that of an electric quadrupole. From the electrical current point of view, the incident electric field induces an oscillating dipole current along the *x*-direction on the top metal patch. Mirrored by the ground plane, there exists a virtual dipole current along the -*x*-direction below the ground plane, with $2t_{LC}$ as the separation between two dipoles. Due to the deep subwavelength thickness of the liquid crystal layer, two dipoles interfere destructively along the $z$-direction ($\theta_{Rx} = 0°$). Dipoles also have radiation nulls along their oscillation directions, which are $-x$ and $+x$-directions ($\theta_{Rx} = -90°$ and $90°$). Therefore, radiation nulls of the proposed structure are at $\theta_{Rx} = -90°, 0°,$ and $90°$, respectively. With the presence of nulls in the unit cell radiation pattern, there is a high loss to form beams towards these angles, consequently degrading the beam-forming performance. The radiation pattern can be improved by redesigning the unit cell to exhibit nulls in $\theta_{Rx} = -90°$ and $90°$ only, resembling that from an electric or magnetic dipole.

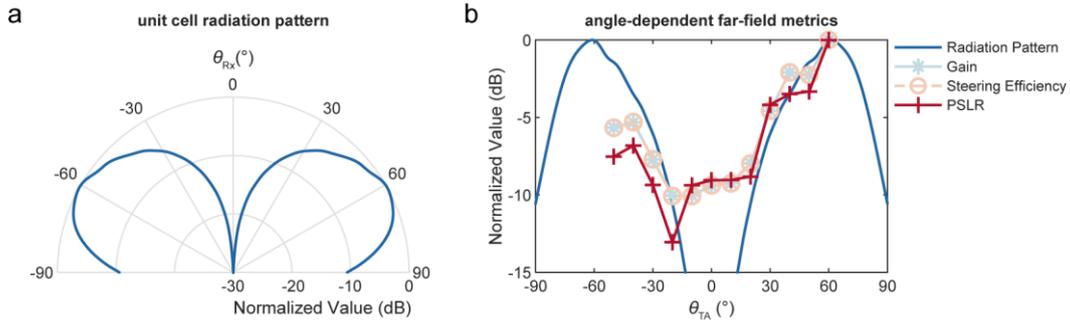

**Figure S4 | Unit cell radiation pattern and far-field metrics at various target angles. a**, The polar plot of unit cell radiation pattern. **b**, Radiation pattern, gain, steering efficiency, and PSLR as a function of $\theta_{TA}$.

### Section 5: Bandwidth and beamwidth at various target angles

In the single-beam-forming measurement result, normalized reflectance can be plotted versus far-field angle $\theta_{Rx}$ and frequency, as shown in Fig. S5a. The frequency-dependent target angle (also known as beam squint) matches the theoretical prediction (dashed lines), showing the dispersive nature of the metasurface. From this reflectance plot, when a fixed $\theta_{TA}$ is selected, the 3-dB bandwidth at each $\theta_{TA}$ can be extracted, as shown in Fig. S5b. The bandwidth decreases with decreasing $\theta_{TA}$, with the bandwidth wider than 9 GHz across the whole FOV.

From this plot, when a fixed $f_{OP}$ is selected, the 3-dB beamwidth at each $\theta_{TA}$ can be extracted, as shown in Fig. S5c. The Beamwidth first decreases and then increases as $\theta_{TA}$ increases, with a minimum around $\theta_{TA} = 0°$, which matches the theoretical beamwidth (dashed line). Because the beamwidth is inversely proportional to the effective area of the metasurface, it is the narrowest at $\theta_{TA} = 0°$ and widens as $\theta_{TA}$ moves away from $0°$. The obtained bandwidth and beamwidth at each $\theta_{TA}$ are also tabulated in Table S1.

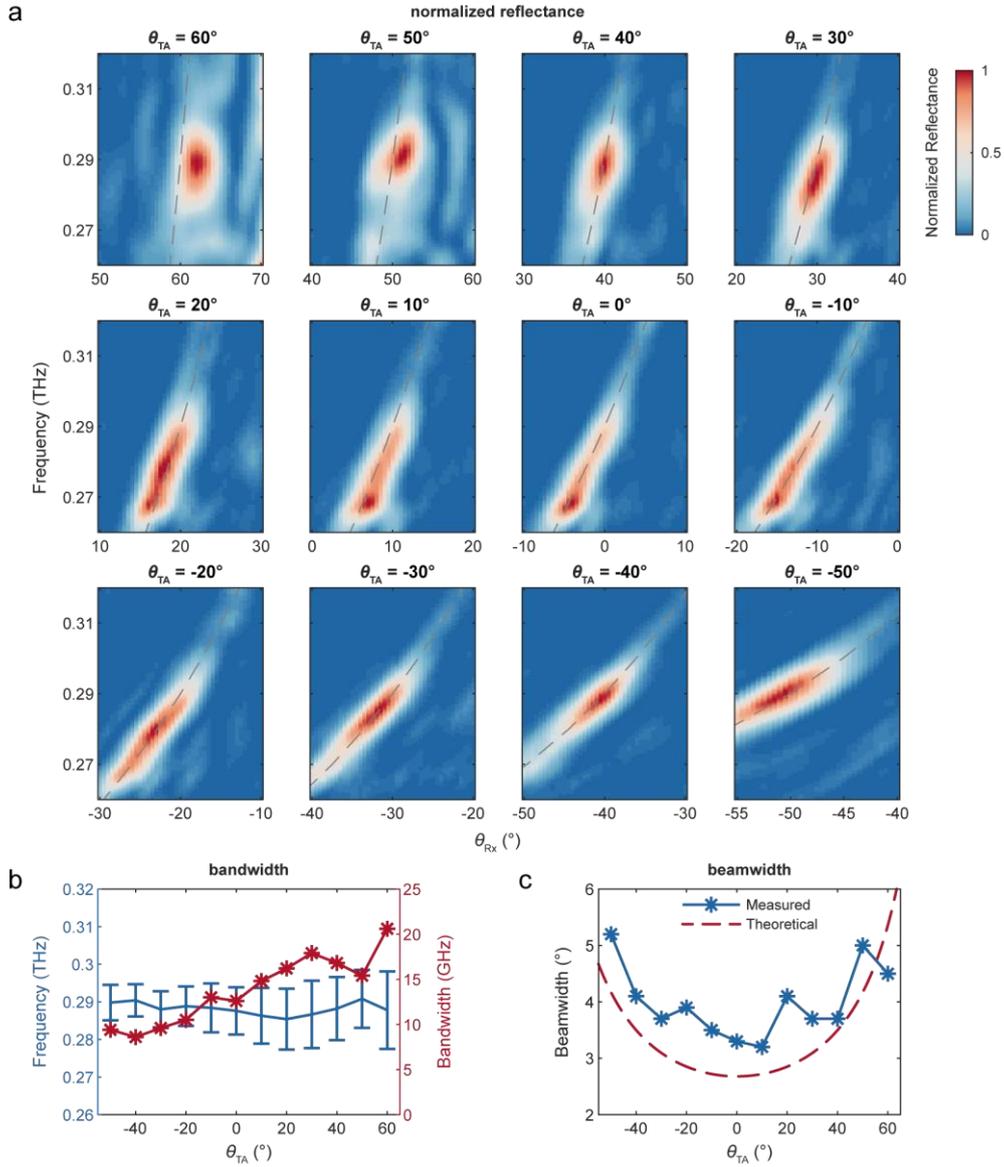

**Figure S5 | Normalized reflectance plot; bandwidth and beamwidth extracted from the plot. a**, Measured normalized angle and frequency-dependent reflectance in the single-beam-forming experiment. The target angle dispersions follow the theoretical predictions (overlayed grey dashed lines). **b**, Measured and theoretical bandwidth as a function of the target angle. The metasurface bandwidth in the whole FOV is wider than 9 GHz. **c**, Measured and theoretical beamwidth as a function of the target angle.

## Section 6: Single-beam-forming with small angular resolution

Besides FOV coverage, the angular resolution of the formed beam is another important metric. Together, these two metrics determine the total number of resolvable angles. To further explore the minimum beam-forming angular resolution, optimization is performed at a series of closely spaced target angles within $\theta_{Rx} = 60° \sim 50°$ and $\theta_{Rx} = -40° \sim -50°$. From the gain patterns in Fig. S6b, the minimum resolution is measured to be 3°, which is comparable to the average beam width. This implies the number of resolvable angles across the 110° FOV is about 40, which may potentially allow user tracking in a finer grain.

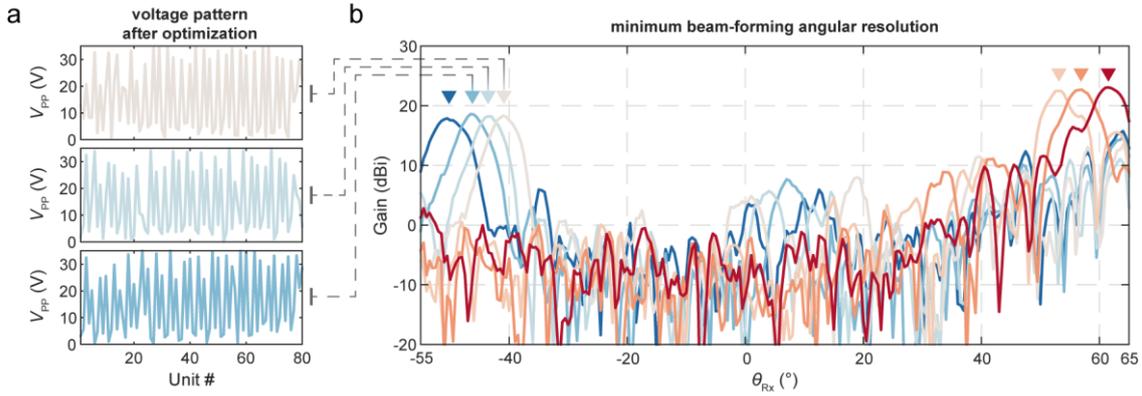

**Figure S6 | Minimum angular resolution of the beam-forming metasurface.** Multiple target angles are selected in an FOV range $\theta_{Rx} = 60° \sim 50°$ and $\theta_{Rx} = -40° \sim -50°$. **a**, Optimized voltage patterns for $\theta_{TA} = -41°, -44°,$ and $-47°$, respectively. **b**, Measured gain patterns after applying optimized voltage patterns for several target angles, indicated by triangular marks, showing that the minimum beam-forming angular resolution is about 3°.

**Section 7: Gain ratio tuning for multi-beam-forming**

For the multi-beam-forming, the relative gain of different beams can be tuned by adjusting the corresponding weight $w_i$ for each beam in a redefined total *FoM*: $FoM_{total} = \Sigma_i w_i FoM_i$, where $FoM_i$ is the *FoM* measured at the $i^{th}$ target angle and $w_i$ is the corresponding weight. Figure S7 illustrates an example of applying weight ratios of $w_1 : w_2 = 1 : 1, 1 : 2,$ and $1 : 3$, respectively, at target angles of $\theta_{TA} = 40°$ & 30°. The relative gain between the two beams is adjusted accordingly.

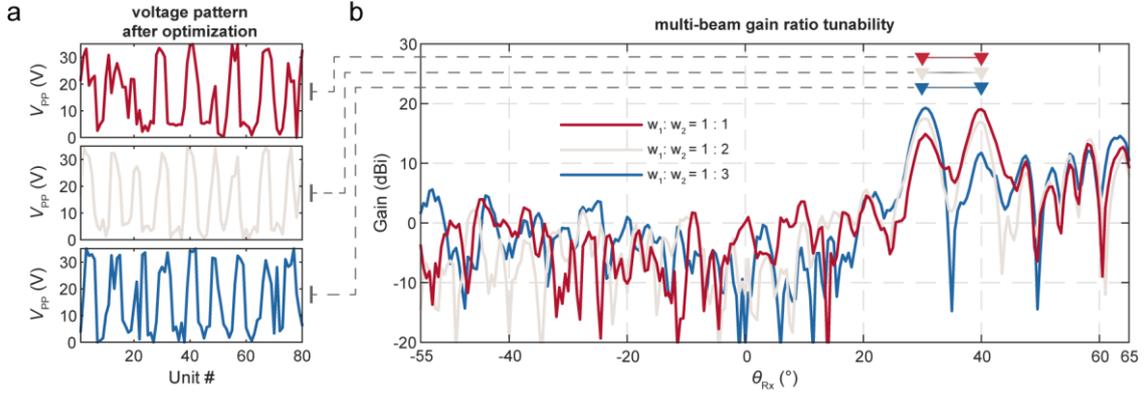

**Figure S7 | Measured beam-forming gain pattern under various weight ratios in the *FoM*, showing the multi-beam gain ratio tunability. a**, Voltage patterns for $\theta_{TA}$ = 40° & 30°, optimized under weight ratios of $w_1 : w_2$ = 1 : 1, 1 : 2, and 1 : 3, respectively. **b**, Measured gain patterns for $\theta_{TA}$ = 40° & 30°, after applying optimized voltage patterns under corresponding weight ratios.

**Section 8: Additional results on metasurface operation with various incident angles**

When performing system networking, the relative position and orientation among multiple metasurfaces should have wide adjustable ranges to ease the deployment. Here, we show that with the beam incident angle $\theta_{Tx}$ varying from -75° to 75°, the metasurface can form beams in a wide range of target angles, as shown in Fig. S8.

In the first case (Fig. S8a-b), the incident angle is varied while fixing the voltage pattern. Formed beams shift away from defined target angles, which showcase a simple way of performing beam steering by shifting the feeding direction. In the second case (Fig. S8c-d), the incident angle is varied and the voltage pattern is re-optimized under the new incident angle. Formed beams can stay at corresponding target angles under varied incident angles. In the third case (Fig. S8e-f), the incident angle is set to $\theta_{Tx}$ = 0° (normal incidence), and the voltage pattern is re-optimized under this incident angle. Target angles of the formed beam can span a wide FOV from $\theta_{TA}$ = 20° to 80°, with progressively wider beamwidth. Considering the symmetry condition, the incident angle can also vary from $\theta_{Tx}$ = 0° to 75°. Since the metasurface can operate within such a wide range of incident angles, its posture in a beam-forming network can be widely adjusted to adapt to various communication scenarios.

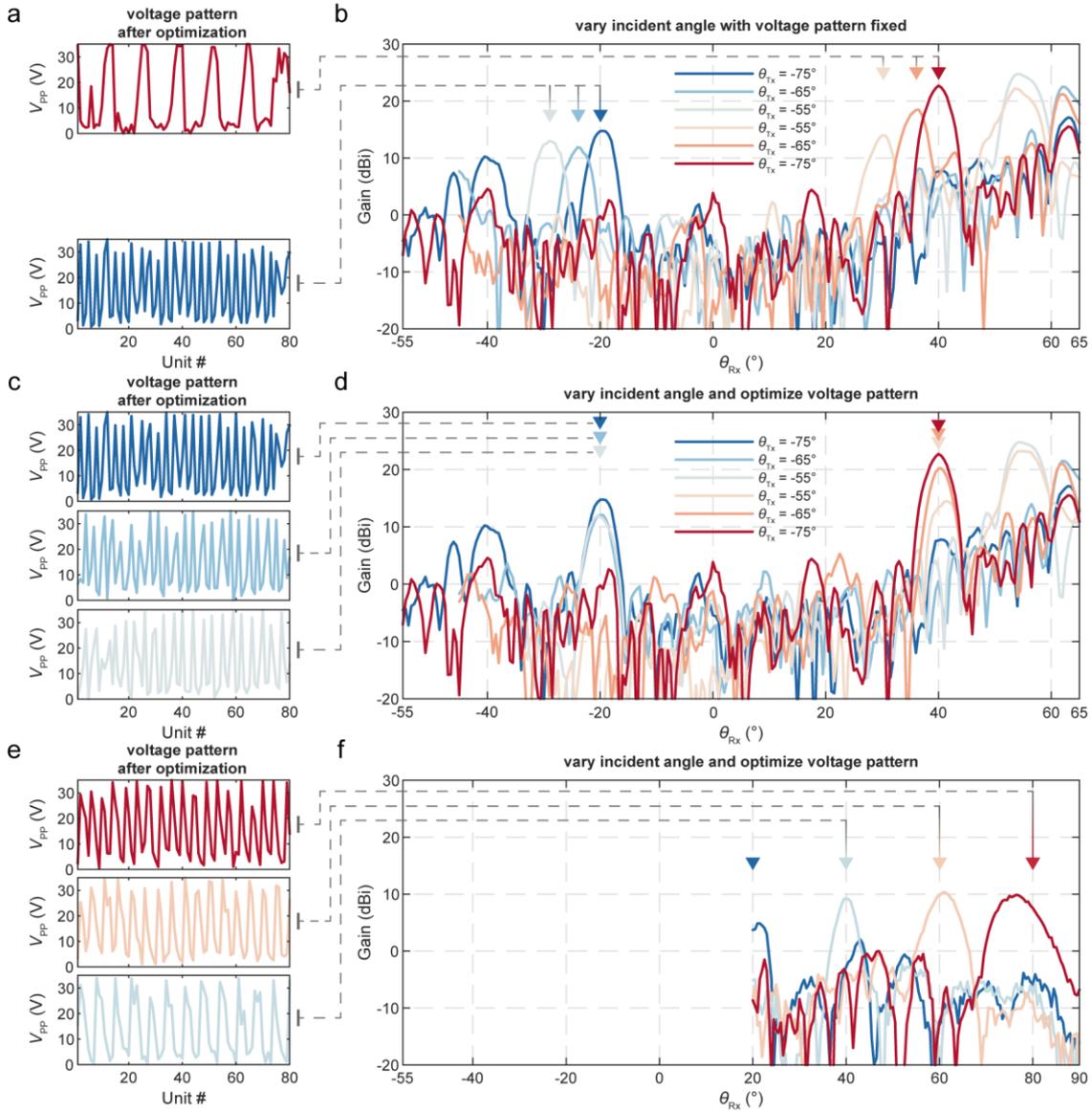

**Figure S8 | Beam-forming under various incident angles, showing the wide incident angle acceptance of the metasurface. a,** Voltage patterns for $\theta_{TA} = 40°$ and $-20°$, optimized under $\theta_{Tx} = -75°$. **b,** Measured gain patterns for $\theta_{TA} = 40°$ and $-20°$ under various incident angles $\theta_{Tx} = -75°$, $-65°$, and $-55°$, respectively. **c,** Voltage patterns for $\theta_{TA} = -20°$, optimized under $\theta_{Tx} = -75°$, $-65°$, and $-55°$, respectively. **d,** Measured gain patterns for $\theta_{TA} = 40°$ and $-20°$ under various incident angles $\theta_{Tx} = -75°$, $-65°$, and $-55°$, respectively, after applying optimized voltage patterns at corresponding incident angles. **e,** Voltage patterns for $\theta_{TA} = 80°$, $60°$ and $40°$, respectively, optimized under $\theta_{Tx} = 0°$. **f,** Measured gain patterns for $\theta_{TA} = 20°$, $40°$, $60°$, and $80°$, respectively, under $\theta_{Tx} = 0°$.

**Section 9: Power consumption measurement**

For the measurement of the power consumption of the metasurface, because the impedance across the liquid crystal layer is high, the current flowing through a single unit is very small. To increase the measurement accuracy, 20 units are bundled together and supplied with a single output from the FPGA board. After applying a maximum 35V $V_{PP}$ square waveform, the total root-mean-square current is measured by an AC current meter to be 80 µA. Consequently, the maximum power consumption of the whole metasurface is:

$$P_{max} = I_{max,rms} V_{max,rms} = \frac{80}{20} \times 80 \text{ µA} \times 17.5 \text{ V} = 5.6 \text{ mW} \tag{S2}$$

**Section 10: Switching speed measurement**

The switching speed of the metasurface is measured using a fiber-based continuous-wave terahertz system (TeraScan 1550, TOPTICA Photonics AG). The experimental setup is similar to that illustrated in Fig. 3a, only replacing the Tx and Rx from the THz-TDS system with the Tx and Rx from the continuous-wave terahertz system. By applying alternating 0V and 35V $V_{PP}$ square waveforms to the metasurface, the power received at Rx alternates between high and low levels, as shown in Fig. S9. Switching times are obtained by measuring the 10%-to-90% rising and 90%-to-10% falling time of the received power. The rising time $t_{raise}$ and falling time are measured to be 210 ms and 640 ms, respectively. The estimated switching speed is therefore 1 s / max ($t_{raise}$, $t_{fall}$) = 1 s / 0.64 s $\approx$ 1.6 Hz.

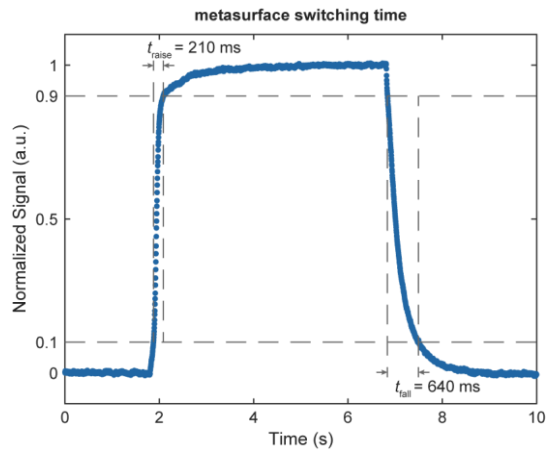

**Figure S9 | Switching time measurement curves and obtained transition time.** The 10% to 90% rising time is measured to be 210 ms, and the 90% to 10% falling time is measured to be 640 ms. The estimated switching speed is 1.6 Hz.

**Section 11: Beam-forming performance under binary voltage mode**

With continuous phase tuning, the voltage pattern optimization procedure is quite time-consuming. During each iteration, several voltage patterns are uploaded to the metasurface. In addition, in our experiments using a THz-TDS system, the delay line in the TDS system is scanned several times for each iteration. For instance, a total of 800 iterations takes about 3 hours.

To accelerate the optimization process, we also test the metasurface under binary voltage mode, in which each unit cell is applied a voltage of either 0 V or 35 V. The voltage value at a unit cell can be determined by binarizing the required phase value: $V_{PP}(m) = 17.5(\text{sign}(\varphi(m)-\pi)+1)$, where $\varphi(m)$ is the desired phase shift at the unit cell $m$. Through this formula, a binary voltage pattern for a specified target angle can be quickly generated without the optimization procedure, as shown in Fig. S10a. The gain pattern measurements from binary voltage patterns are shown in Fig. S10b.

A detailed comparison of beam-forming metrics under various operation modes is shown in Fig. S10c. From this comparison, binary voltage patterns result in lower performances than the optimized continuous voltage patterns but are still better than the theoretical voltage patterns without considering mutual coupling between neighboring unit cells. Thus, binary voltage patterns can be decent approximations of the optimized continuous voltage patterns, although optimization is still required to achieve a better beam-forming performance, including a further increased gain, steering efficiency, and PSLR.

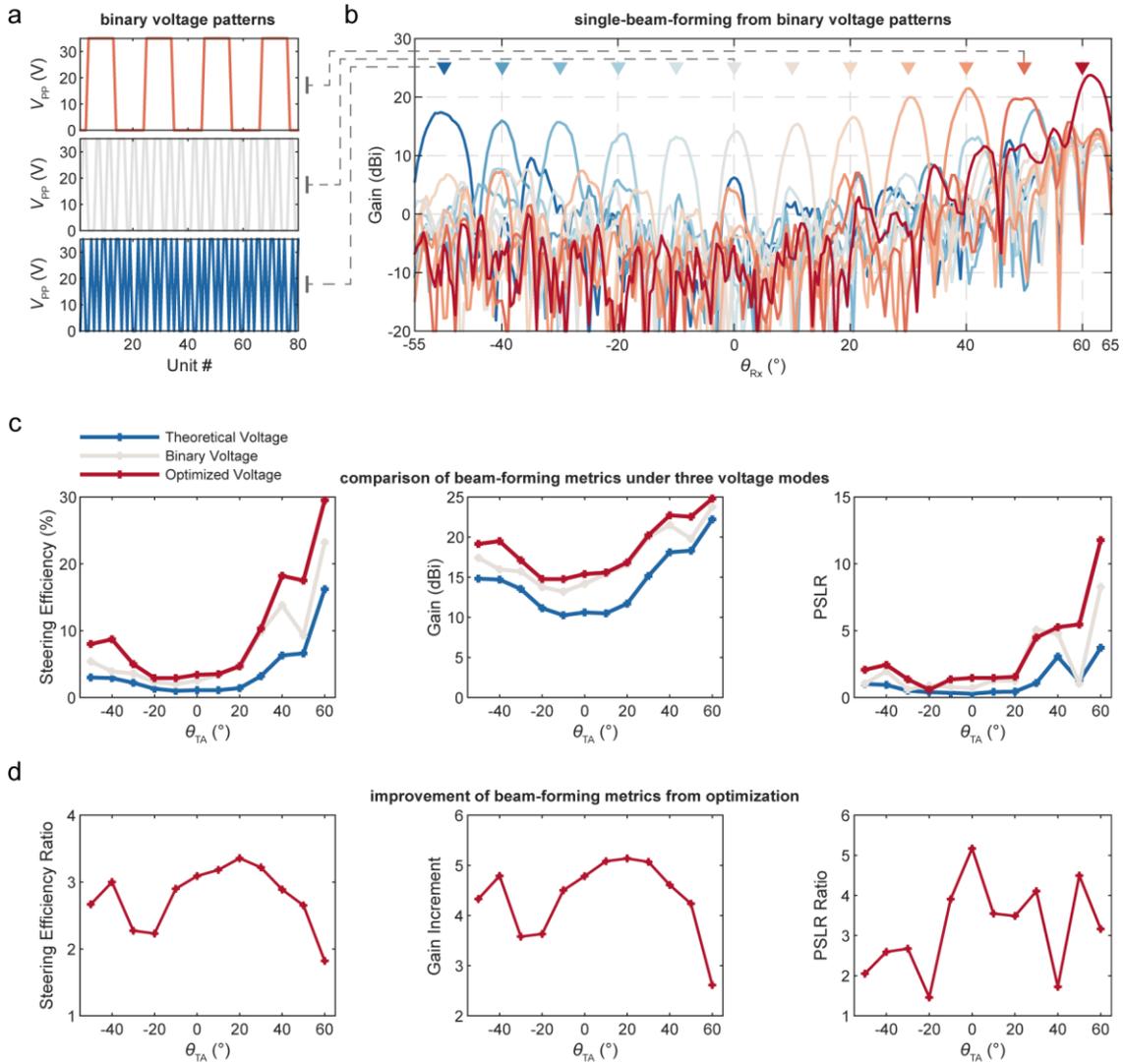

**Figure S10 | Measured gain patterns from binary voltage patterns and beam-forming performance comparison under various voltage modes. a**, Binary voltage patterns for target angles $\theta_{TA}$ = 50°, 0°, and -50°, respectively. **b**, Measured gain patterns after applying binary voltage patterns for 12 target angles, indicated by triangular marks. **c**, Comparison of beam-forming metrics (gain, steering efficiency, and PSLR, respectively) under three voltage modes: theoretical voltage converted from the desired phase, binary voltage, and optimized voltage, respectively. **d**, Improvement of beam-forming metrics from the optimization process, comparing beam-forming metrics from optimized voltage patterns with theoretical voltage patterns.